\documentclass[doublecol,english]{epl2}
\usepackage{graphics}
\usepackage{amsmath}

\usepackage{graphicx}
\usepackage{ifsym}
\usepackage{wasysym}
\usepackage{babel}
\usepackage{siunitx}

\title{Wrinkling instability of vesicles in steady linear flow}

\author{Michael Levant\inst{1}, David Abreu\inst{2}, Udo Seifert\inst{2}, and Victor Steinberg\inst{1}}

\institute{
\inst{1} Department of Physics of Complex Systems, Weizmann Institute of Science, Rehovot, 76100 Israel \\
\inst{2} II. Institut f\"ur Theoretische Physik, Universit\"at Stuttgart, 70550 Stuttgart, Germany}

\date{\today}

\abstract{We present experimental observations and numerical simulations of a wrinkling instability that occurs at sufficiently high strain rates in the trembling regime of vesicle dynamics in steady linear flow. Spectral and statistical analysis of the data shows similarities and differences with the wrinkling instability observed earlier for vesicles in transient elongation flow. The critical relevance of thermal fluctuations for this phenomenon is revealed by a simple model using coupled Langevin equations that reproduces the experimental observations quite well.}

\pacs{87.16.D-}{Membranes, bilayers, and vesicles}
\pacs{47.63.-b}{Biological fluid dynamics}
\pacs{05.40.-a}{Fluctuation phenomena, random processes, noise, and Brownian motion}

\begin{document}
\maketitle

\section{Introduction}
Wrinkling is a well-known phenomenon frequently observed in nature and everyday life on widely different spatial scales \cite{cerda,cerda1}. It bridges mechanics, geometry, physics, and biology. It usually occurs either in extensible materials due to external stretching, or as a result of compression of inextensible materials, in which case it reduces to the Euler buckling instability \cite{landau}. In contrast to the wrinkling instability in films and structures such as micro-capsules, where it occurs due to stretching elasticity \cite{fink06}, in vesicles with unstretchable lipid membranes it takes place under compression due to negative surface tension \cite{vasiliywrinkles,turitsynwrinkles}.

\begin{figure}
\centering
  \includegraphics[width =0.6\linewidth]{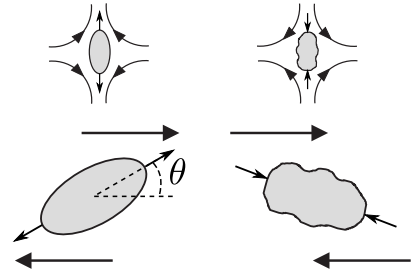}
  \caption{Schematic presentation of flow reversal and vesicle compression in elongation (top) and linear (bottom) flows.}
  \label{fig:4roll-mill}
\end{figure}

Vesicles are fluid droplets encapsulated by a phospholipid-bilayer membrane and suspended in either the same or another fluid \cite{seifertreview}. They are considered as simple model objects to simulate red-blood-cell dynamics in a first step towards understanding blood rheology. Two constraints, namely the conservation of both vesicle volume $V$ and surface area $A$, result in rather involved non-linear dynamics in flow (for a review, see \cite{abre14}). In transient plane elongation (hyperbolic) flow, the wrinkling instability happening during the relaxation of a vesicle towards a new stationary state was studied experimentally \cite{vasiliywrinkles}. There, the flow direction was reversed faster than the vesicle's relaxation time scale $\tau\sim\eta_\text{out}R_0^3/\kappa$, where $\kappa$ is the bending elasticity, $\eta_\text{out}$ outer fluid viscosity, and $R_0=(3V/4\pi)^{1/3}$ the effective vesicle radius defined via the volume $V$. This sudden reversal causes a switching from vesicle stretching to vesicle compression, see Fig. \ref{fig:4roll-mill} top. The next reversal then takes place after full relaxation. For strain rates $\chi$ lower than a critical value $\chi_c$, this flow reversal leads to an almost smooth transition from an elliptical vesicle with long axis in the direction of compression to one with long axis perpendicular to it. During this transient transformation, only deformations with wave number $k=2$ are induced by the flow. Higher-order deformations are only excited thermally, and the deformation power spectrum $P_k\sim k^{-4}$ is observed since bending deformations enter as $\kappa k^4$ in the membrane energy  \cite{seifertreview}. For $\chi>\chi_c$, higher-order deformation modes, coined wrinkles, are generated on top of an almost elliptical vesicle. A spectral analysis of the shape distortions easily identifies the wrinkling onset due to a sharp growth of higher-order harmonics (see Fig. 3 in \cite{vasiliywrinkles}).

   \begin{figure*}
\centering
  \includegraphics[width = 0.8\textwidth]{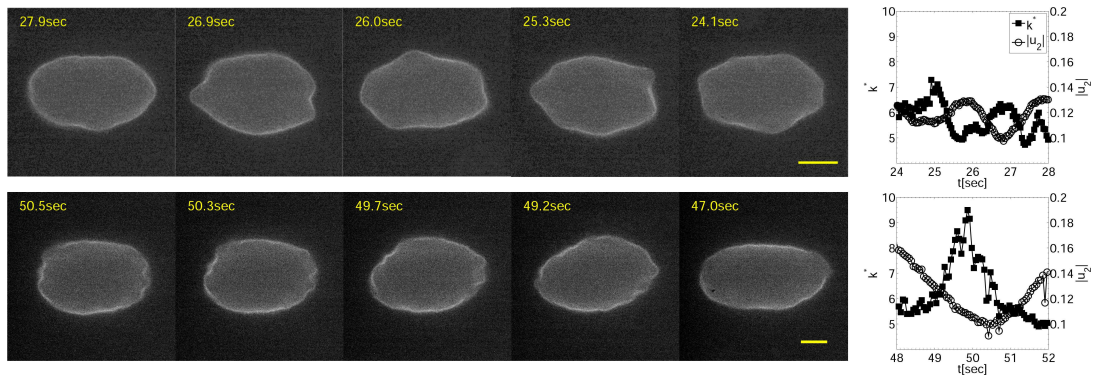}
  \caption{Snapshots of vesicle shape perturbations at $S=421$, $\Lambda=1.87$, $\Delta=1.22$ (top, see movie1.avi (1 MB)) and $S=1147$, $\Lambda=1.78$, $\Delta=1.27$ (bottom, see movie2.avi (1 MB)). Right: corresponding time series of $k^*$ and $|u_2|$. Scale bar $=10\mu m$.}
  \label{fig:snaps}
\end{figure*}
In this Letter, we explore another mechanism to observe wrinkling caused by the interplay of thermal fluctuations with the trembling (TR) motion. The TR regime of vesicle dynamics in steady linear flow is an intermediate regime between tank-treading (TT) and tumbling (TU) \cite{kantsler2}. It is characterized by quasi-periodic noisy oscillations of the vesicle main axis around the shear direction, i.e., oscillations of the inclination angle $\theta$ around zero. The lower images of Fig. \ref{fig:4roll-mill} illustrate schematically the flow inversion and vesicle compression in linear flow. The flow direction makes an angle of $\pi/4$ with the direction of the extensional part of the flow. At negative inclination angles $\theta<0$, a reversal of the elongation part of the flow occurs in the vesicle frame of reference. Then, the vesicle is compressed for some part of its oscillation period, leading to wrinkling at sufficiently large values of the normalized strain rate $S=14\pi s\eta_\text{out}R_0^3/(3\sqrt{3}\kappa\Delta)$ \cite{lebedev}, where $s$ is the strain rate of the flow, $A$ the vesicle area, and $\Delta=A/R_0^2-4\pi$ the excess area.

In spite of the fact that a possible connection between vesicle shape deformations and a wrinkling-like instability in TR was discussed already in Refs. \cite{julien1,julien2,norman}, only the recent observation of the amplification of thermal noise by TR dynamics \cite{mynoise,davidnoise} provides a real basis to this assumption. Indeed, in spite of a small ratio of thermal energy to bending rigidity $k_BT/\kappa \approx 0.05$, where $k_B$ is the Boltzmann constant and $T$ the temperature, the dramatic effect of thermal noise on vesicle dynamics subjected to a linear flow was recently observed in experiment \cite{mynoise}. Numerical simulations including thermal fluctuations reproduce the experimental observations on a quantitative level \cite{davidnoise}. Both works such as previous stochastic simulations \cite{noguchi,messlinger} show that thermal noise is the main factor for the strong discrepancies in vesicle dynamics found for the last several years between experiment and various numerical simulations based on deterministic models. The most prominent discrepancies are the difference in vesicle dynamics between TR in experiment \cite{kantsler2} and vacillating-breathing (VB) in deterministic simulations \cite{misbah}, and the number of the control parameters to present the phase diagram of the vesicle dynamical states in linear flow \cite{julien1,norman,farutin}. The experimental results show that two parameters, which represent the competition between relevant time scales, are enough to describe the transition between TT, TR, and TU, namely $S$ and $\Lambda=4(1+23\lambda/32)\sqrt{\Delta}\omega/s$, where $\lambda=\eta_\text{in}/\eta_\text{out}$ is the viscosity contrast ($\lambda=1$ in this article) and $\omega$ the flow vorticity \cite{lebedev}.

The inclusion of thermal noise in the model leads to strongly distorted vesicle shapes due to appearance of higher-order harmonics, in particularly odd ones, which become larger than the second harmonic during the compression at negative $\theta$ (Figs. 2 in \cite{mynoise,davidnoise}). As a result, at large values of $S$ a wrinkling instability similar to that observed in a transient elongation flow could show up\cite{vasiliywrinkles}. Since the TR period is not well defined and vesicle compression and stretching occur continuously during each cycle, the vesicle shape perturbations can propagate between each flow reversal into the next cycle, additionally contributing to the irregularities of the TR motion and the wrinkling instability. Thus, the quantitative analysis of wrinkling in TR is much less straightforward than in elongation flow, since wrinkles appear on the top of several even and odd modes with sufficiently large amplitudes that continuously vary during a cycle. Furthermore, the power spectrum below the wrinkling instability is very different from the thermal noise spectrum. All these factors hinder an unambiguous identification of the wrinkling onset and scaling above it in TR.

In this Letter we present experimental results that clarify the following questions: (i) Is the wrinkling instability possible in TR? (ii) What is the control parameter? (iii) Is it possible to quantitatively characterize the wrinkling transition and scaling above it in the TR regime? Moreover, we present analytical and numerical results which reproduce the main characteristics of the wrinkling instability and show that thermal noise plays a crucial role.

\section{Experimental results}
In order to quantitatively investigate the wrinkling instability and the vesicle shape deformations in TR in a wide range of $S\in[10;2400]$, which was chosen as the control parameter, $\Lambda\in[1.3;2.42]$, and $\Delta\in[0.43;1.65]$,  we modified our experimental setup (see e.g. \cite{mynoise}), making it suitable to higher-viscosity solvents by decreasing the resistance of the inlets, and increased the accessible flow rates. As a result, $S$ was increased by more than one order of magnitude while the Reynolds number remained of the order of $10^{-4}$.

\begin{figure}
\centering
  \includegraphics[width =0.9\linewidth]{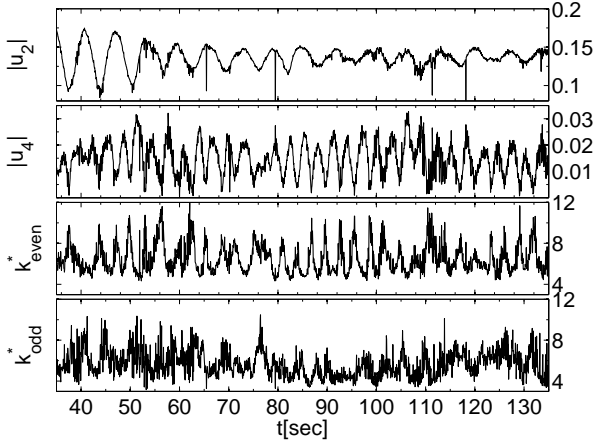}
  \caption{Dynamics of $|u_2|$, $|u_4|$, $k^*_\text{even}$, and $k^*_\text{odd}$ at $S=1147$ ($S_\text{eff}=757$), $\Lambda=1.78$, and $\Delta=1.27$.}
  \label{fig:k_star}
\end{figure}
We observed 14 vesicles, using $\eta_\text{out}=\eta_w$, $5.3\eta_w$, and $10.66\eta_w$, where $\eta_w=\SI{1}{\milli\pascal\second}$ is the viscosity of water. Observations and measurements of vesicle dynamics were conducted inside a micro-fluidic 4-roll mill, implemented in silicone elastomer (Sylgard 184, Dow Corning), fabricated via soft lithography \cite{julien2,muller,mynoise}. The key component of this device is a dynamical trap, that allows long time observation of vesicles in a planar linear flow (observation times were long compared to the timescales of the flow). The flow inside the trap is generated using hydrostatic pressure. The physical control parameter of the flow inside the trap, namely the ratio of vorticity to strain rate $\omega/s$ inside the trap, is controlled by the single experimental control parameter, the pressure drop across the dynamical trap that can be varied continuously. The flow rates were measured by particle tracking velocimetry \cite{grier}.
\begin{figure}
\centering
  \includegraphics[width = 0.8\linewidth]{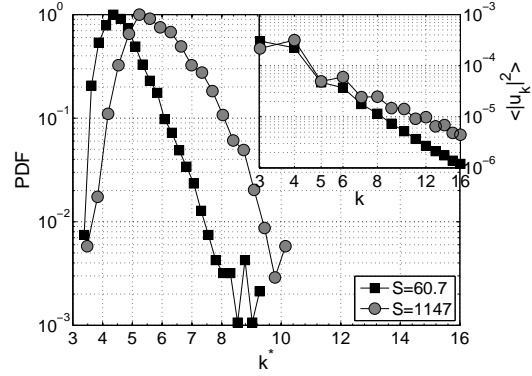}
  \caption{PDF of $k^*$ on a log-linear scale showing exponential decay for sufficiently long time series. The maxima of the distributions are set to $1$ for better comparison. Inset: Power spectra $|u_k|^2$ versus $k$ show amplification of the higher-order modes for larger S. The vesicle parameters are $S=60.7$, $\Lambda=1.8$, $\Delta=0.66$ and $S=1147$, $\Lambda=1.78$, $\Delta=1.27$. }
  \label{fig:k_star_hist}
\end{figure}
Vesicles were prepared in water, $38\% w/w$ sucrose/water and $60\% w/w$ glycerol/water solutions ($1,5.3$ and $\SI{10.66}{\milli\pascal\second}$, respectively, $\lambda=1$), using electroformation method \cite{angelova}, as described in \cite{mynoise}. The dynamics of the vesicles was monitored using inverted fluorescence microscope (IMT-2, Olympus). The images of the vesicle at its largest cross section were collected with a Prosilica EC1380 CCD camera, aligned with the shear axis and synchronized with a mechanical chopper on the path of the excitation beam to reduce unnecessary exposure time. The frame rate/shutter time were adjusted to match the characteristic timescales that vary with the flow rates. Further experimental details were reported elsewhere \cite{mynoise}.

The relative position of the membrane $r(\phi,t)$ was determined in the frame of reference of the vesicle using intensity variations along the radial directions. For each image we obtained up to 500 discrete positions sampled along the contour. To quantitatively analyze the TR dynamics the dimensionless shape of the vesicle $u(\phi,t)=r(\phi,t)/R(t)-1$, $0\leq\phi\leq2\pi$, $R(t)=\langle r(\phi,t)\rangle_\phi$, was Fourier-decomposed via fast Fourier transform, i.e. $u(\phi,t)$ was expressed as $u(\phi,t)=\sum_k{u_k(t)e^{-ik\phi}}$. Then the vesicle shape perturbations were quantified via a mean wave number $k^*=\sqrt{\sum_{k=3}^{16}{k^2|u_k|^2}/\sum_k{|u_k|^2}}$, similarly to the analysis done in transient elongation flow \cite{vasiliywrinkles}.

Figure \ref{fig:snaps} presents snapshots of two time series of TR at two values of $S= 421$ and $S=1147$, much higher than reported before. One can easily identify, besides the previously observed concavities, wrinkles with significantly higher wave number modes. On the right hand side, the time series of the second mode $|u_2|$ and $k^*$ corresponding to the same time intervals show their anti-correlation - the attenuation of $|u_2|$ due to compression is compensated by the amplification of higher-order modes. However, during much longer time intervals $k^*$ exhibits irregular temporal dynamics, weakly correlated with the dynamics of the second mode. Therefore, we defined $k^*_\text{even}$ and $k^*_\text{odd}$, where summation over the even and the odd modes in the expression for $k^*$ was applied. We found that $k^*_\text{even}$ shows quasi-regular dynamics with clearly visible peaks anti-correlated with $|u_4|$ and with about the same period, which is about half that of $|u_2|$, whereas the dynamics of $k^*_\text{odd}$ is rather erratic (Fig. \ref{fig:k_star}). This observation can be explained by the fact that the symmetric even modes are associated with the dynamics, whereas the odd modes are associated with the amplification of thermal noise, as discussed in \cite{mynoise}.

\begin{figure}
\centering
  \includegraphics[width = 0.6\columnwidth]{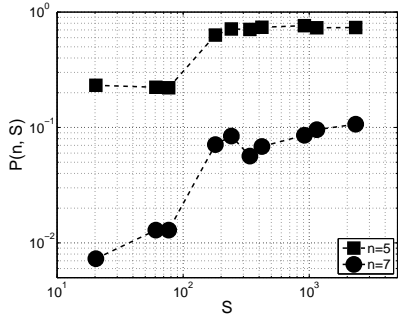}
  \caption{The normalized probability ${\cal P}(n,S)$ as function of $S$, at $n=5$ (squares) and $n=7$ (circles) for all $\Lambda$ and $\Delta$ used in the experiment. }
  \label{fig:probs3}
\end{figure}

\begin{figure}
\centering
  \includegraphics[width = 0.6\columnwidth]{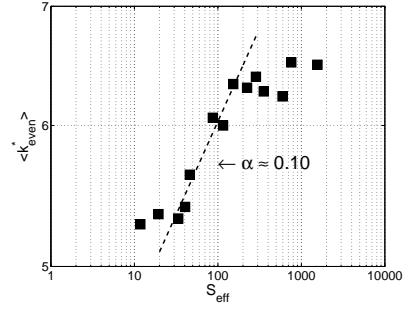}
  \caption{$\langle k^*_\text{even}\rangle$ versus $S_\text{eff}$ for all $\Lambda$ and $\Delta$.}
  \label{fig:transition}
\end{figure}

Another way to analyze the shape perturbations is to present the probability distribution function (PDF) of $k^*$  and time averaged power spectra of the vesicle shape at different $S$. It is obvious from both plots in Fig. \ref{fig:k_star_hist} that the higher $S$, the higher the probability to excite higher harmonics and the larger the spectral amplitude of the higher-order modes. The quantity $k^*$ is a reliable quantitative indication of the wrinkle generation at larger $S$. Due to strong interference of the amplified thermal noise with the wrinkling, we use further a probabilistic approach to characterize the wrinkling onset. We calculate the probabilities of $k^*$ to be above a certain mode number $n$ as function of $S$ that is calculated as the ratio of the PDF area at $k^*>n$ normalized by the total PDF area according to ${\cal P}(n,S)\equiv P(k^*>n,S)/P(3<k^*<16,S)$. As one can see in Fig. \ref{fig:probs3}, ${\cal P}(n,S)$ increases up to four times between $S\simeq 80$ and $S\simeq 150$ for $n=5$ and $n=7$ for all $\Lambda$ and $\Delta$ used in the experiment.

An approach similar to one used in \cite{vasiliywrinkles,julien1} can be also applied. In this case, we modify the control parameter to be $S\rightarrow S_\text{eff}=S\cos(\theta_{rms}+\pi/4)$ that is rotated by $\pi/4$ with respect to the compression axis (see Fig. \ref{fig:4roll-mill}). Then the mean value $\langle k^*_\text{even}\rangle$  shows a transition as a function of $S_\text{eff}$  independent of $\Lambda$ and $\Delta$ used in the experiment. It scales as $S_\text{eff}^{0.10}$ in the interval $S_\text{eff}\in[40;120]$, and is independent of $S_\text{eff}$ outside of this interval (Fig. \ref{fig:transition}). This behavior is in qualitative agreement with the behavior of $k^*$ as function of the normalized strain in the transient elongation flow reported in \cite{vasiliywrinkles}.

\section{Theoretical analysis}
From a theoretical perspective, the main mechanism triggering the wrinkling instability can be understood from the model for quasi-spherical vesicles in flow presented in \cite{seifert}. The three-dimensional radius $r(\theta,\phi)$ of a vesicle is expanded in spherical harmonics ${\cal Y}_{lm}$ as
\begin{equation}
	r(\theta,\phi)=R_0(1+\sqrt{\Delta}\sum u_{lm}{\cal Y}_{lm}(\theta,\phi))\equiv R_0(1+u)
\end{equation}
where the sum goes over $l>1$ and $|m|\leq l$, $\Delta$ is the excess area, and $R_0$ the effective radius. This vesicle is subject to the linear flow
\begin{equation}
	{\bf v}^\infty=\omega(y{\bf e_x}-x{\bf e_y})+s(y{\bf e_x}+x{\bf e_y})
\end{equation}
of vorticity $\omega$ and strain rate $s$. Then the equation of motion for the single modes $u_{lm}$ is
\begin{equation}
	\tau\partial_t u_{lm}=i\frac{S\Lambda}{2}u_{lm}-\mu_l(l(l+1)+\sigma)u_{lm}- i\frac{S}{2}f_{lm} + \zeta_{lm}
	\label{eq_ulm}
\end{equation}
with $\tau\equiv(385/72)\sqrt{\pi/10\Delta}\eta_\text{out}R_0^3/\kappa$, $f_{lm}\equiv \pm\delta_{l,2}\delta_{m,\pm2}$, and $\mu_l\equiv(l-1)l(l+1)(l+2)\tau\kappa/((4l^3+6l^2-1)\eta_oR_0^3)$. The homogeneous surface tension $\sigma$ is a dynamical parameter which is adjusted such that the vesicle area is kept constant. The correlations of the Gaussian white noise $\zeta_{lm}$ are given by the fluctuation-dissipation theorem in equilibrium.

We first neglect thermal noise and decompose the expansion coefficients as $u_{lm} \equiv r_{lm}\exp(-im\theta_{lm})$. We obtain from \eqref{eq_ulm} the equation of motion
\begin{equation}
	\partial_t r_{lm}=-\mu_l(l(l+1)+\sigma)r_{lm}+ \frac{S}{2}r_{22}\sin(2\theta_{22})f_{lm}
	\label{eq_rlm}
\end{equation}
for the amplitudes $r_{lm}$. For $(l,m)\neq(2,\pm 2)$, $f_{lm}=0$ and Eq. \eqref{eq_rlm} describes an exponential decay or growth, depending on the sign of $l(l+1)+\sigma$. The surface tension $\sigma$ can be obtained from the condition that the excess area $\Delta$ remains constant over time, i.e, that the equality
\begin{equation}
	1=\sum_{l,m}\frac{(l-1)(l+2)}{2}r_{lm}^2
	\label{eq_del}
\end{equation}
is always obeyed \cite{seifert}. Taking the time-derivative of Eq. \eqref{eq_del} leads to
\begin{align}
	&l(l+1)+\sigma= \nonumber \\
	&\frac{4Sr_{22}\sin(2\theta_{22})+\sum_{l',m'}[l(l+1)-l'(l'+1)]\nu_{l'}r_{l'm'}^2}{\sum_{l',m'}\nu_{l'}r_{l'm'}^2}
	\label{eq_sig}
\end{align}
with $\nu_l\equiv(l-1)(l+2)\mu_l$. Whether the mode $(l,m)$ decays or grows depends on the sign of \eqref{eq_sig}. We first see that positive contributions come from all modes with $l'<l$: the excess area is transferred from high-order modes to low-order ones. Moreover, the external flow has a contribution which has the sign of the angle $\theta_{22}$, which is approximately the inclination angle, and a strength proportional to $S$. In TR and TU motions, the flow compresses the vesicle when the inclination angle takes negative values, leading to negative surface tension and amplification of the modes with $l(l+1)+\sigma<0$. With increasing $S$, higher-order modes will be excited during compression phases since the surface tension will take larger negative values. However, harmonics with $l>2$ obeying Eq. \eqref{eq_rlm} will decay to $0$ in the long run due to the stretching at positive inclination angles and the transfer of excess area from high to low-order harmonics.

In  Eq. \eqref{eq_ulm}, the term $iSf_{lm}/2$ is derived from the quantities $X^\infty\equiv{\bf v^\infty}(r=R_0)\cdot{\bf e_r}$ and $Y^\infty\equiv-R_0\nabla\cdot{\bf v^\infty}(r=R_0)$ which describe the flow at the virtual sphere $r=R_0$ \cite{seifert}. If we instead define $X^\infty$ and $Y^\infty$ at $r=R_0(1+u)$, correction terms of the form $Su_{lm}$ will induce a coupling between $u_{2,\pm2}$ and higher-order modes with even $l$ and $m$. However, modes with odd $l$ and/or $m$ will still vanish. Therefore, we have to take thermal noise into account to explain the presence of odd modes in experiments. In the following, we present simulation results of Eq. \eqref{eq_ulm} with thermal noise ($T=\SI{293}{\kelvin}$, $\kappa=25k_BT$) and the mentioned additional flow correction terms. To enforce the constraint \eqref{eq_del}, we replace $\sigma$ by a quadratic energy term with large spring constant \cite{davidnoise}. We simulate trajectories of several thousand periods such that initial transient effects vanish. In order to compare the simulations with the experiments, the shape in the middle flow plane is expanded in Fourier coefficients $u_k$ like in the experiments.

\begin{figure}
\centering
  \includegraphics[width = 0.80\columnwidth]{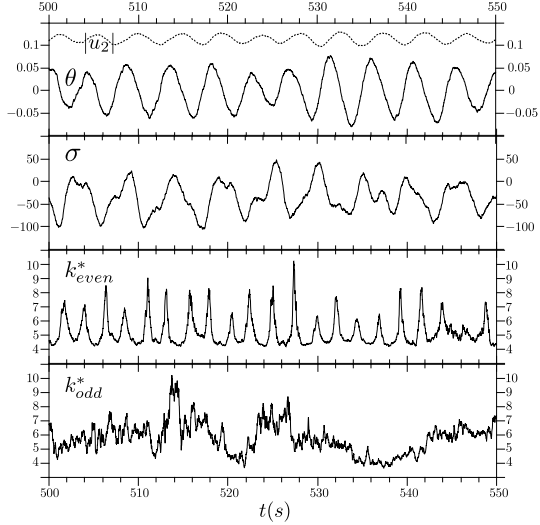}
  \caption{Simulation results for the dynamics of the inclination angle $\theta$ (full line) and the second mode amplitude $|u_2|$ (dashed), the surface tension $\sigma$, and the mean wave numbers $k^*_\text{even}$ and $k^*_\text{odd}$ for $S=1147$, $\Lambda=1.78$, and $\Delta=1.27$.}
  \label{fig:sim1}
\end{figure}
All simulations with the same physical parameters as in Fig. \ref{fig:transition} predict a TR motion in the steady state, in agreement with experiments. For instance, Fig. \ref{fig:sim1} shows a simulated trajectory for $S=1147$, $\Lambda=1.78$, and $\Delta=1.27$ corresponding to Fig. \ref{fig:snaps} bottom. The first panel shows the almost regular oscillation of the inclination angle $\theta$ around $0$. The amplitude of the second mode $|u_2|$ oscillates with the same period, which is comparable to the experimental one (see Fig. \ref{fig:snaps}), although it lags slightly behind. The surface tension $\sigma$ shown in the second panel, which corresponds to Eq. \eqref{eq_sig} with flow correction terms, is strongly correlated with $\theta$, reaching minima and maxima almost at the same times. This strong correlation is the basis of the wrinkling phenomenon. The third and fourth panel show $k^*_\text{even}$ and $k^*_\text{odd}$, respectively. As in the experimental data of Fig. \ref{fig:k_star}, $k^*_\text{odd}$ exhibits a rather erratic evolution while $k^*_\text{even}$ is regular since even modes are coupled by the flow.  Moreover, like in the experiments, $k^*_\text{even}$ has a period approximately equal to half the period of $\theta$. This fact is due to thermal fluctuations, as explained below.

\begin{figure}
\centering
  \includegraphics[width = 0.80\columnwidth]{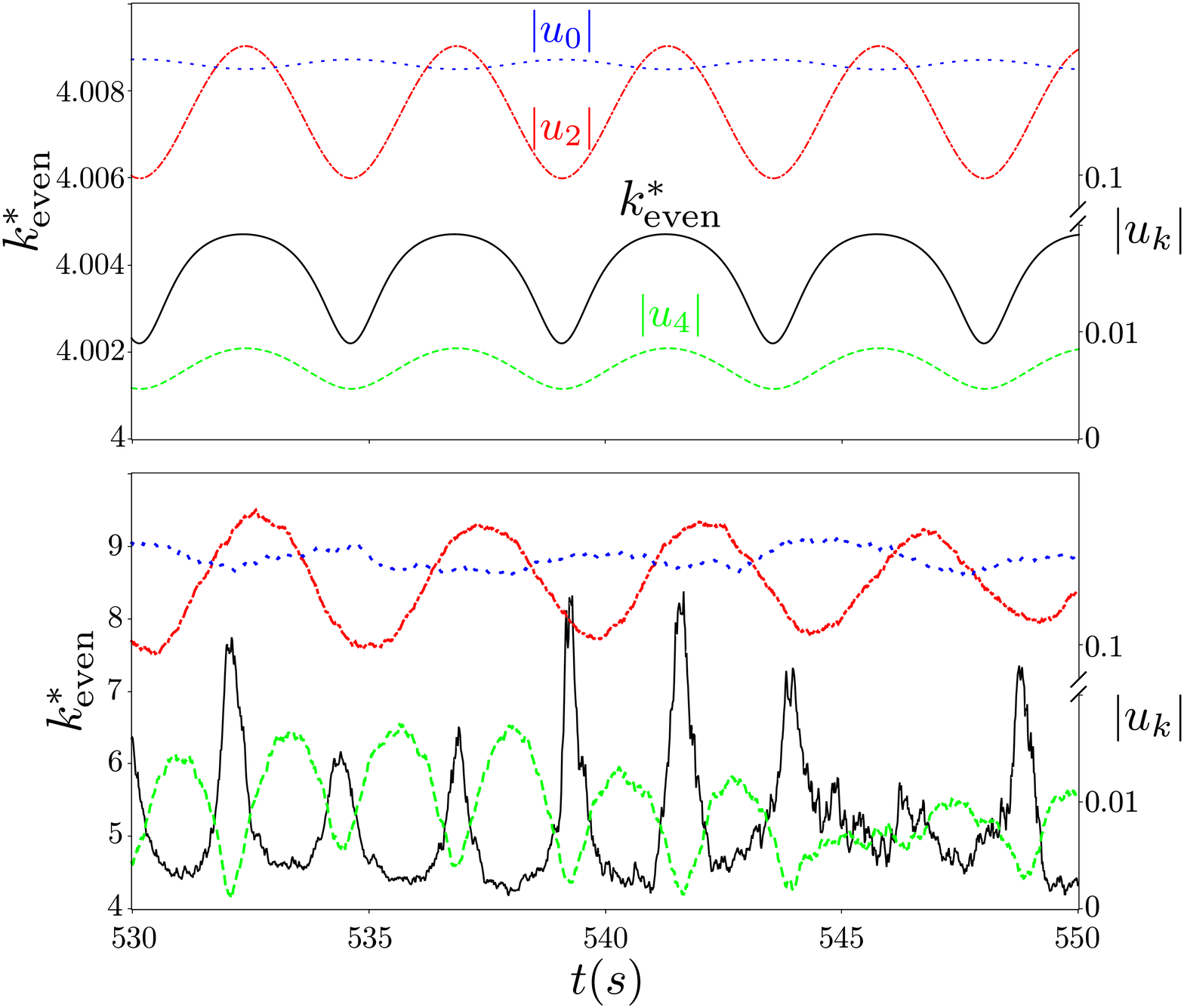}
  \caption{Evolution of $|u_0|\equiv R(t)/R_0-1$ (blue dots), $|u_2|$ (red dash-dots), $|u_4|$ (green dashes), and $k^*_\text{even}$ (black) without (top) and with (bottom) thermal noise. The left scale is for $k^*_\text{even}$ and the right one (which has a break) for the $|u_k|$.}
  \label{fig:sim2}
\end{figure}

Fig. \ref{fig:sim2} compares the evolution of the zeroth, second, and fourth modes with that of $k^*_\text{even}$. The top panel shows the deterministic case. It corresponds to a vacillating-breathing motion \cite{misbah}: when the inclination angle is negative, the amplitude of the second mode $|u_2|$ decreases while $|u_0|$ and modes out of plane increase (see also data in Ref. \cite{norman}). The dynamics of all higher-order even modes, e.g., $|u_4|$, is perfectly synchronized with $|u_2|$, as is the evolution of $k^*_\text{even}$. In addition, $k^*_\text{even}$ varies between $4.002$ and $4.005$, meaning that the fourth mode is always much larger than the higher ones. With thermal fluctuations, the picture changes dramatically. The zeroth and second modes do not show qualitative changes, but $|u_4|$ has a totally different behavior. It now exhibits a maximum not only when $|u_2|$ is maximal but also when it is minimal. This new peak is due to the amplification of thermal noise at negative inclination angles \cite{mynoise,davidnoise} and corresponds to the observed wrinkles. Moreover, the mean wave number $k^*_\text{even}$ reaches large values, comparable with the experimental ones, and is strongly anti-correlated with $|u_4|$. The explanation is that when the inclination angle reaches its minimum, high-order deformations grow fast, thus increasing $k^*_\text{even}$. However, as Eq. \eqref{eq_sig} shows, this excess area is transferred to low-order deformations, leading to a decrease of $k^*_\text{even}$ and a growth of $|u_4|$, which promptly decreases again in favor of $|u_2|$. Then the inclination angle becomes positive, and $|u_4|$ increases again due to the deterministic coupling with $|u_2|$. Therefore, thermal fluctuations not only excite odd modes, but change drastically the dynamics of the even modes as well.

\begin{figure}
\centering
  \includegraphics[width = 0.9\columnwidth]{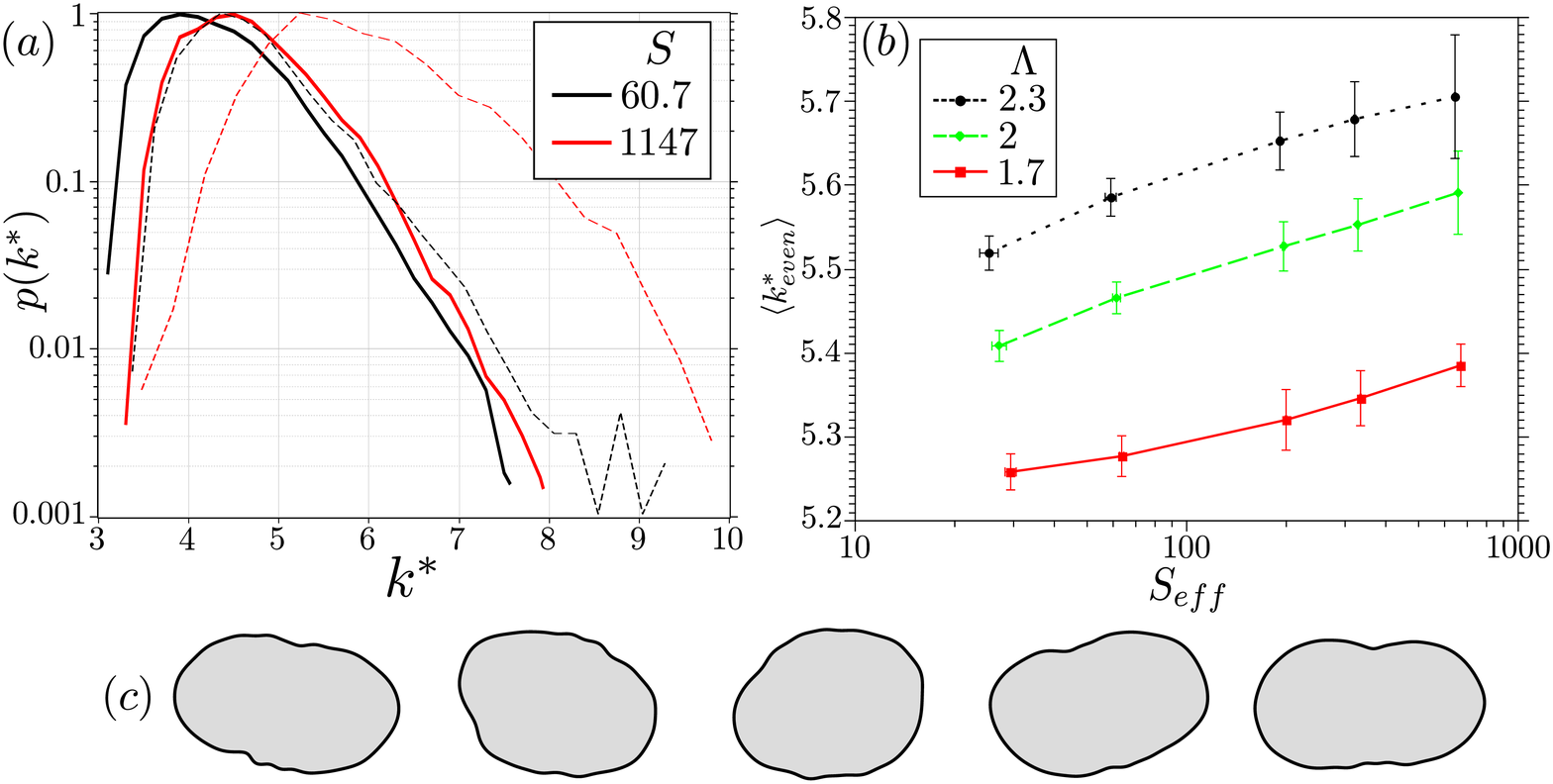}
  \caption{(a) PDF of $k^*$ for $S=60.7$, $\Lambda=1.8, \Delta=0.66$ and  $S=1147$, $\Lambda=1.78, \Delta=1.27$ in simulations (full lines) and experiments (dashed lines, from Fig. \ref{fig:probs3}). (b) Mean $k^*_\text{even}$ as a function of $S_\text{eff}$ for $\Delta=1$ and different values of $\Lambda$. (c) Snapshots from simulations at $S=421$, $\Lambda=1.87, \Delta=1.22$.}
  \label{fig:sim3}
\end{figure}
Fig. \ref{fig:sim3} shows simulation results that can be compared to experimental ones. The probability distribution function of $k^*$ of Fig. \ref{fig:sim3}(a) shows the same main features as Fig. \ref{fig:probs3}: for  $S=1147$, it has a larger tail and a higher average than for $S=60.7$, even though the difference is not as large as in the experiments. Furthermore, the simulations could not reproduce the transition shown in Fig. \ref{fig:transition} due to a marked dependence not only on $S$ but also on $\Lambda$ and $\Delta$. Fig. \ref{fig:sim3}(b) shows, however, that for fixed $\Delta$ and $\Lambda$, the expected increase in the average $k^*_\text{even}$ is observed, although with a smaller magnitude than in experiments. The snapshots at $S=421$ of Fig. \ref{fig:sim3}(c) show clearly visible wrinkles which are, however, not as pronounced as the ones of Fig. \ref{fig:snaps}. The numerical results seem to underestimate the experimental ones. This discrepancy is probably due to the quasi-spherical approximation which does not describe faithfully vesicles with large $\Delta$. Higher-order curvature terms in the bending energy might also trigger short-wavelength instabilities of the membrane \cite{franke}. However, in spite of these approximations, the model still captures most experimental features, mainly because it includes thermal fluctuations consistently.

\section{Conclusion}
We have shown that a wrinkling instability occurs for vesicles in the TR regime at large enough strain rates. Even though the dynamics is much more complex than in pure elongation flow, the experiments show that the normalized strain rate $S$ is the main control parameter triggering the instability. The onset of wrinkling occurs approximately above $S\simeq60$, which comes from a statistical analysis of the power spectrum. The main experimental features are reproduced by a simple numerical model, which shows that thermal fluctuations are crucial to describe wrinkling correctly. This model does not, however, provide a full quantitative agreement with the experiments due to the quasi-spherical approximation. In order to compare theory and experiment on a more quantitative level, it would be interesting to include thermal noise in more complex numerical schemes such as \cite{zhao11,yazd12}, which show a strong coupling between even modes at large $S$, or to analyze single TR vesicles with the algorithm of \cite{mcwh09}, which already takes thermal fluctuations into account.

\acknowledgments
 This work is partially supported by a grant from the German-Israeli Foundation.

\bibliographystyle{eplbib}
\bibliography{wrinklingdynamics2}

\end{document}